\documentclass[twocolumn,aps,tightenlines]{revtex4}

\usepackage{amsmath}
\usepackage{amssymb}
\usepackage{bbm}
\usepackage{simplewick}
\usepackage{mathtools}
\usepackage[normalem]{ulem}
\usepackage{dsfont}
\usepackage{mathrsfs}
\usepackage[makeroom]{cancel}
\usepackage{graphicx}

\def\x{\boldsymbol{x}}
\def\y{{\boldsymbol{y}}}
\def\z{{\boldsymbol z}}

\def\0{\boldsymbol{0}}

\def\J{\boldsymbol{J}}

\def\tj{{\tilde J}}
\def\tJ{\boldsymbol{\tj}}

\def\p{{\boldsymbol{p}}}

\def\k{{\boldsymbol{k}}}

\def\GB{{GB}}
\def\pGB{{|\psi_\GB\rangle}}
\def\HGB{{{\cal H}_\GB}}

\def\rl{{\rangle}{\langle}}

\def\II{{{\Rzymskie{2}}}}

\def\sigtr{\text{$\sigma=1,2$}}
\def\sigtrPrim{\text{$\sigma'=1,2$}}

\def\HZERO{{{\cal H}_0}}

\def\A{\boldsymbol{A}}
\def\E{\boldsymbol{E}}

\def\Bold{\boldsymbol{B}}

\def\T{\mathds{T}}

\def\Lag{{\cal L}}

\def\ii{\mathrm{i}}

\def\d#1{d^3\mkern-1.5mu#1\,}

\def\dd#1{d^4\mkern-1.5mu#1\,}

\def\ddd#1{\frac{d^3\mkern-1.5mu#1}{(2\pi)^3}}
\def\dddd#1{\frac{d^4\mkern-1.5mu#1}{(2\pi)^4}}

\def\bnabla{{\boldsymbol\nabla}}

\def\beps{{\boldsymbol\epsilon}}

\newcommand{\rfock}[3]{
\ifthenelse{\equal{#2}{}}{|#1,\lfloor#3\rfloor\ra}{|#1,\lceil#2\rceil,\lfloor#3\rfloor\ra}
}
\newcommand{\lfock}[3]{
\ifthenelse{\equal{#2}{}}{\la#1,\lfloor#3\rfloor|}{\la#1,\lceil#2\rceil,\lfloor#3\rfloor|}
}

\def\om#1{\omega_{\boldsymbol{#1}}}

\def\la{\langle}
\def\ra{\rangle}

\def\cl{\text{cl}}

\def\xymunu{(x,\mu\leftrightarrow y,\nu)}

\def\B#1{\!\left(#1\right)}
\def\BB#1{\!\left[#1\right]}

\def\N#1{:\!#1\!:}

\def\be{\begin{equation}}
\def\ee{\end{equation}}

\def\bee{\begin{equation*}}
\def\eee{\end{equation*}}

\def\bg{\begin{equation}\begin{gathered}}
\def\eg{\end{gathered}\end{equation}}

\def\bgg{\begin{equation*}\begin{gathered}}
\def\egg{\end{gathered}\end{equation*}}

\def\ba{\begin{equation}\begin{aligned}}
\def\ea{\end{aligned}\end{equation}}

\def\izero{\ii0}

\def\DIV{\text{div}}

\def\spp{\text{spin}}
\def\fl{\text{field}}

\def\orbb{\text{orb}}

\def\for{\ \text{for} \ }

\makeatletter
\newcommand{\pushright}[1]{\ifmeasuring@#1\else\omit\hfill$\displaystyle#1$\fi\ignorespaces}
\newcommand{\pushleft}[1]{\ifmeasuring@#1\else\omit$\displaystyle#1$\hfill\fi\ignorespaces}
\makeatother

\def\XXint#1#2#3{{\setbox0=\hbox{$#1{#2#3}{\int}$}
\vcenter{\hbox{$#2#3$}}\kern-.5\wd0}}

\newcommand{\Rzymskie}[1]{%
  \textup{\uppercase\expandafter{\romannumeral#1}}%
}

\usepackage[ddmmyyyy]{datetime}
\makeatletter
\def\Dated@name{}
\makeatother

\begin{document}
\title{Impact  of  gauge fixing  on   angular momentum 
operators of the covariantly quantized electromagnetic field}
\author{Bogdan Damski}
\affiliation{Jagiellonian University, Institute of Theoretical Physics, {\L}ojasiewicza 11, 30-348 Krak\'ow, Poland}
\begin{abstract}
Covariant quantization of the electromagnetic field imposes  the so-called
gauge-fixing modification on the Lagrangian density. As a result of that,
the total angular momentum operator receives at least one gauge-fixing-originated
contribution, whose presence causes some confusion in the literature. 
The  goal  of this work is to discuss
in detail why such a contribution, having no 
 classical interpretation, is actually indispensable.
For this purpose, we divide   canonical and  Belinfante-Rosenfeld
total angular momentum operators into different components and study 
their commutation relations, their role in generation of rotations of quantum fields, 
and their action on states from the physical sector of the theory.
Then, we examine  physical matrix elements of operators having 
gauge-fixing-related  contributions,
illustrating  problems that one may  encounter due to 
careless employment of the resolution of identity 
during their evaluation. 
The resolution of identity, 
in the indefinite-metric space of 
the covariantly-quantized electromagnetic field,
is extensively discussed  because it takes a not-so-intuitive form if one insists 
on explicit projection onto states from the physical sector of the theory.
Our studies are carried out in  the framework of 
the Gupta-Bleuler theory of the free electromagnetic field. 
Relevant remarks 
about  interacting systems, described  by covariantly-quantized electrodynamics, are  given.
\end{abstract}
\maketitle

\section{Introduction}
\label{Introduction_sec}
We all know that covariantly-quantized electrodynamics has some 
appealing features. One is that its  free-field photon propagator 
has the simplest possible form, which facilitates perturbative
calculations. The other is that its  
electromagnetic vector potential  $A^\mu$
is manifestly Lorentz covariant, which comes in handy   when 
relativistic transformation properties  of  operators 
and matrix elements involving $A^\mu$  are   discussed.
In addition to that, such $A^\mu$ is local, which  is  
important in certain rigorous studies  of
quantum field theory 
(see e.g. Sec. 1.1 of \cite{KugoProg1979} for  relevant references).
These and related convenient features of the covariant quantization 
framework do not come for
free. The price one has to pay for them can be seen in different ways.

To begin, one has to accept  the fact that unphysical
degrees of freedom are  being introduced by the quantization procedure, 
scalar and longitudinal photons, critically commented upon 
by Schwinger \cite{SchwingerPR1959}, Strocchi  \cite{StrocchiPR1970},
and others. 
To make matters worse, the    abnormal commutation 
relation of  creation and annihilation operators
of 
scalar photons leads to the conclusion  that  $A^\mu$ 
is defined on the 
indefinite-metric space \cite{Nagy,Nakanishi1972}, where vectors can have 
positive, negative, or zero norm \cite{remarkNorm}. 

This result asks for the definition of the physically-relevant  sector of such a space,
and for this purpose one typically assumes that matrix elements  of 
$\partial\cdot A$
vanish in  states from such a subspace. The very existence of such a condition 
strikingly illustrates problems with enforcement of
the Lorenz gauge on the operator level, which can
be seen as a one more complication.
These rather unwelcome features 
were  technically addressed in early papers of Gupta \cite{GuptaPRSA1950}  and Bleuler \cite{BleulerHelv1950}, 
quantizing the electromagnetic field in the   Feynman gauge
(see \cite{Gupta} for the comprehensive summary of these efforts).
Those studies  were subsequently generalized to 
account for other  covariant gauges--especially  the
Landau gauge--by Lautrup \cite{Lautrup1967} and  Nakanishi \cite{Nakanishi1972}
(see  Sec. 18 of \cite{Nakanishi1972} for  comments about \cite{Lautrup1967},
 Sec. 7 of \cite{Greiner} for  textbook discussion of \cite{Lautrup1967}, and
 \cite{GieresNPB2021} for a recent overview).

As far as this work is concerned, the main technical complication 
of the covariant quantization 
approach is that it modifies the Lagrangian density by adding the so-called 
gauge-fixing term to it. 
The most  elementary  reason for doing so is that 
it solves the problem
of vanishing momentum canonically conjugate to $A^0$. This 
allows for quantization of all components of the electromagnetic vector potential,
which is of key importance in the covariant quantization framework.
The gauge-fixing modification of the Lagrangian density can be 
also seen as a Lagrange multiplier, imposing the Landau gauge  
in the appropriate limit (see e.g. Sec. 15.5 of \cite{WeinbergII}). 

Such a procedure leads to  field  equations that do not correspond
to  Maxwell (Maxwell-Dirac) 
equations in  free (interacting) systems.
Interestingly,  
this complication can be, in the free theory
of the electromagnetic field, 
rigorously linked  to 
locality  and Lorentz covariance of $A^\mu$ (these properties are necessarily
lost when Maxwell's equations are kept as operator identities
\cite{StrocchiPR1967,StrocchiPR1970}). 

The change of the Lagrangian density automatically affects various  
observables derived from it. This remark brings our attention 
to  conserved quantities in general and 
angular momentum, being of special interest here, in particular.
Namely, the total angular momentum operator  receives at least one gauge-fixing 
correction, which causes some confusion.

The confusion arises when one 
(i) starts from the classical theory, which neither
needs  nor has any gauge-fixing modifications;
(ii) derives the classical expression for  total
angular momentum; 
(iii) replaces the classical electromagnetic vector potential in it 
by the covariantly-quantized one; (iv) assumes that the resulting expression 
represents the total angular momentum operator.

In general,  such a procedure  leads to the expression violating   some of the 
most basic properties expected of the total angular momentum operator.
Its failure  can be traced back to the  (iii) step, 
which is done under the tacit assumption that the covariant quantization can
be  carried out without the gauge-fixing modification of the Lagrangian
density. 
This is not the case, which we have already mentioned. 
To fix the above procedure, it is sufficient to 
alter  the (i) step 
by working from the very beginning with the modified  Lagrangian density,
inevitably introducing    gauge-fixing angular  momentum to the theory.
The  purpose of this work is to discuss 
in detail why such an awkward kind of angular momentum, completely redundant in the
classical theory, is actually indispensable on the quantum level.

This can be done in the most transparent manner  in the free theory of the
electromagnetic field, which we will be exploring 
in the following sections 
(relevant remarks about the interacting fields will be also given).
Our  discussion will be focused on    two most popular  ways  in which 
the   (ii) step is performed: the canonical and 
Belinfante-Rosenfeld decompositions.

These and other decompositions, frequently commented upon in literature
\cite{LeaderPhysRep2014short,WakamatsuIntJModA2014},
aim at division  of the total angular momentum operator into 
physically-meaningful contributions.
The following question then arises. What is the
role of  the gauge-fixing angular momentum operator in such considerations? 

Given  substantial 
interest in  angular momentum-related physics, it is a bit surprising that 
this issue is poorly explored in the literature. For example, nothing can be 
found about the gauge-fixing angular momentum operator in 
popular review \cite{WakamatsuIntJModA2014} and numerous other
works tacitly assuming that the gauge-fixing procedure is absent, 
while at the same time intensely  discussing the division of the 
total angular momentum operator into gauge-invariant components \cite{remarkGaugeFixingWakamatsu}.
These considerations do {\it not} apply to 
covariantly-quantized theories, such as the 
Feynman gauge one that we study, because   they
are carried out irrespective of the quantization procedure necessarily 
bringing the gauge-fixing angular momentum operator to the  table.

The  inescapable fact  of  its presence, in  
studies assuming that the operator  $A^\mu$ transforms as a genuine $4$-vector,
was   stated  in \cite{LeaderPRD2011}. In particular, 
one can find  in this article interesting  critical 
assessment of papers  
ignoring gauge-fixing-originated 
contributions to total linear and angular momentum operators.
In this context,  our work 
comprehensively   
illustrates the  importance of the latter contributions, approaching 
the subject from a different perspective than \cite{LeaderPRD2011}.

If one acknowledges  that there are gauge-fixing-originated 
contributions to
various observables in a covariantly-quantized theory, 
then it is of interest to know whether they 
contribute to physical matrix elements (i.e. the ones computed in 
states from the physically-relevant sector  of the theory). 
This issue was discussed in \cite{LeaderPRD2011} for the  density of the energy-momentum tensor. 
It was  argued there that its gauge-fixing-related part has
vanishing physical matrix elements, 
and then this statement  was reiterated in popular review \cite{LeaderPhysRep2014short}. 
However, the argument leading to such a conclusion relies on insertion 
of a complete set of physical states 
between  field operators in the considered  matrix elements \cite{LeaderPRD2011}. 
We question validity of 
such an argument, supporting our view by a detailed discussion of the subtle issues 
associated with the resolution of identity in the indefinite-metric space.

Before presenting the outline of this work, we  
mention that the gauge-fixing modification of the Lagrangian density 
can be avoided when one works in   the non-covariant gauge that is
effected   
on the operator level. For example, this is the case in the 
Coulomb gauge studies,
if one parameterizes   the spatial part of the 
electromagnetic vector potential operator 
so as to satisfy the $\DIV\A=0$ constraint \cite{Greiner}. 
Then  clearly some aspects of angular momentum-related 
studies are simpler. However,  the pleasant
features of the covariant quantization framework, including
the ones
mentioned at the beginning of this section, are lost. This is why we believe
that it is worth to face 
subtleties
of the  covariant approach
head-on. Moreover, we mention that it is 
our opinion that covariant gauge studies of angular momentum
physics are   underrepresented 
in the literature, which additionally motivates us  for pursuing them.
Finally, it should be said that at least some non-covariant 
gauges can be also
implemented by the gauge-fixing modification of the Lagrangian 
density (see e.g. \cite{AdkinsPRD1983} for the Coulomb
gauge example).  Its influence on the total angular momentum operator  
could  be presumably analyzed akin to what we do 
in this work.

The outline of this paper is the following. Sec. \ref{Classical_sec} 
briefly presents  classical results on 
angular momentum of the electromagnetic field   from the 
perspective relevant to our studies.
Sec. \ref{QuantumI_sec} is devoted to quantum investigations of the canonical decomposition of 
the total angular
momentum operator. It starts, however,
with the introduction to the Gupta-Bleuler
quantization scheme, before focusing on  three
components of such an operator.
Their commutation relations, their role in  generation of rotations of
quantum fields, and their action on  states from the physical sector
of the theory are  extensively  discussed in Secs. \ref{QuantumI_secA},
\ref{QuantumI_secB}, and  \ref{QuantumI_secC}, respectively. Analogical analysis
is then presented in Sec.
\ref{QuantumII_sec} for the Belinfante-Rosenfeld decomposition.
Next, physical matrix
elements of  gauge-fixing-related operators
are discussed in 
Sec. \ref{Gauge_sec}. Particular stress is placed there
upon presentation of the correct form of the resolution
of identity  in the indefinite-metric space because 
it takes a not-so-intuitive form
if one insists on explicit projection onto states from the physical sector
of the theory. This discussion is then continued in Appendix \ref{Two_app}. 
The summary of our work is given  in Sec. \ref{Summary_sec}.

\section{Classical considerations}
\label{Classical_sec}
The classical Lagrangian density for the free electromagnetic field is given by  
\be
\Lag^\cl=-\frac{1}{4}F^\cl_{\mu\nu}F_\cl^{\mu\nu},
\label{Lclass}
\ee
where $F^\cl_{\mu\nu}=\partial_\mu A_\nu^\cl-\partial_\nu A_\mu^\cl$ and 
the symbol $\cl$ is used to distinguish  classical expressions 
from 
their quantum counterparts studied in the following sections.
To specify the remaining conventions, we  mention that we 
adopt the Heaviside-Lorentz system of units
and set  $\hbar=c=1$. Moreover, 
we use the metric tensor $\eta=\text{diag}(+---)$ and assume that 
Greek and Latin indices of tensors  take values $0,1,2,3$ and   $1,2,3$,
respectively. The Einstein summation convention is frequently applied to
them. $3$-vectors are written in bold, e.g. $x=(x^\mu)=(x^0,\x)$.
The Levi-Civita symbol is written as $\varepsilon^{ijk}$, where  $\varepsilon^{123}=+1$.

To obtain the canonical expression for angular momentum, one evaluates \cite{Greiner} 
\begin{subequations}
\begin{align}
&J_\cl^i=\frac{1}{2}\varepsilon^{imn}
\int \d{z}(\vartheta_\cl^{0n}z^m -\vartheta_\cl^{0m}z^n +\delta M_\cl^{0mn}),\\
\label{emtensor}
&\vartheta_\cl^{\mu\nu}=\frac{\partial\Lag^\cl}{\partial(\partial_\mu A_\cl^\sigma)}\partial^\nu A_\cl^\sigma
- \eta^{\mu\nu}\Lag^\cl,\\
&\delta M_\cl^{\mu\nu\lambda}= 
\frac{\partial\Lag^\cl}{\partial(\partial_\mu A_\cl^\sigma)}
(I^{\nu\lambda})^{\sigma\rho}A^\cl_\rho,\\
&(I^{\alpha\beta})^{\gamma\delta}=\eta^{\alpha\gamma}\eta^{\beta\delta}-\eta^{\alpha\delta}\eta^{\beta\gamma},
\end{align}
\label{Js}%
\end{subequations}
where $\vartheta_\cl^{\mu\nu}$ stands for the  canonical energy-momentum tensor 
density   while $\delta M_\cl^{\mu\nu\lambda}$ is  the so-called spin 
contribution.
In the end, one finds that (\ref{Js}) has the following spin and
orbital components
\begin{subequations}
\begin{align}
&\J^\cl=\J^\cl_\spp + \J^\cl_\orbb,\\
&\J^\cl_\spp=\int\d{z} \E^\cl\times\A^\cl,\\
&\J^\cl_\orbb=\int\d{z}  E_\cl^j(\z\times\bnabla)A_\cl^j,
\end{align}
\label{canon}%
\end{subequations}
where neither Euler-Lagrange equations nor changes  of the 
integrands through addition or subtraction of $3$-divergence terms 
have been employed  during derivation of (\ref{canon}).

A different procedure for defining angular momentum appears when 
one works with the  $\mu\leftrightarrow\nu$    symmetric   energy-momentum tensor density
\cite{Greiner}
\begin{subequations}
\be
\tilde{\vartheta}_\cl^{\mu\nu}=\vartheta_\cl^{\mu\nu}+\partial_\sigma\chi_\cl^{\sigma\mu\nu},
\ee
\be
\begin{aligned}
\chi_\cl^{\sigma\mu\nu}= \frac{A^\cl_\lambda}{2}&\left[
\frac{\partial\Lag^\cl}{\partial(\partial_\sigma A_\cl^\rho)}(I^{\mu\nu})^{\rho\lambda}
+\frac{\partial\Lag^\cl}{\partial(\partial_\nu A_\cl^\rho)}(I^{\mu\sigma})^{\rho\lambda}\right.\\
&\left.-\frac{\partial\Lag^\cl}{\partial(\partial_\mu A_\cl^\rho)}(I^{\sigma\nu})^{\rho\lambda}
\right],
\end{aligned}
\ee
\label{chi3}%
\end{subequations}
which is obtained from (\ref{chi3})  after employment of Euler-Lagrange equations. It is used
for computation of  angular momentum via 
\be
\tj_\cl^i=\frac{1}{2}\varepsilon^{imn}\int \d{z}
(\tilde{\vartheta}_\cl^{0n}z^m-\tilde{\vartheta}_\cl^{0m}z^n).
\label{tildeJ}
\ee
In the end, one arrives at the Belinfante-Rosenfeld expression \cite{Belinfante1939,Rosenfeld,Belinfante1940}
\be
\tJ^\cl=\int\d{z} \z\times(\E^\cl\times\Bold^\cl).
\label{Jinv}
\ee

Thorough discussion of  quantum equivalents
of  
(\ref{canon}) and (\ref{Jinv})
will be presented in Secs. \ref{QuantumI_sec} and
\ref{QuantumII_sec}, respectively.

\section{Quantum considerations: canonical expressions}
\label{QuantumI_sec}
Upon transition to the  quantum theory, 
the classical  vector potential $A^\mu_\cl$ gets replaced 
by the operator $A^\mu$, which may suggest that 
the total canonical angular momentum operator 
will be given by (\ref{canon}) stripped off the symbol $\cl$. 
It is actually not what is happening, which we have already  commented upon in
Sec. \ref{Introduction_sec}.

It has been also mentioned there 
that the theory based solely on  Lagrangian density (\ref{Lclass})
cannot be covariantly quantized.
This issue is resolved by adding the  gauge-fixing
term to (\ref{Lclass})
\be
\Lag^\cl\to\Lag=-\frac{1}{4}F_{\mu\nu}F^{\mu\nu}-\frac{1}{2}(\partial\cdot A)^2,
\label{Lcov}
\ee
which leads to the following canonical commutation relations \cite{Greiner}
\begin{subequations}
\begin{align}
\label{Acan1}
&[A^\mu(x),A^\nu(y)]=0,\\
&[\partial_0A^\mu(x),A^\nu(y)]= \ii\eta^{\mu\nu}\delta(\x-\y),\\
&[\partial_0A^\mu(x),\partial_0A^\nu(y)]=0,
\end{align}
\label{Acan}%
\end{subequations}
where equal times, $x^0=y^0$, are  assumed \cite{remarkEqtimes}.

The  vector potential operator then  reads
\begin{subequations}
\begin{align}
&A^\mu(z)= \int
\frac{\d{k}}{(2\pi)^{3/2}}
\sum_{\sigma=0}^3
\frac{\epsilon^\mu(\k,\sigma)c_{\k\sigma} }{\sqrt{2\om{k}}}e^{-\ii k\cdot z} 
+\text{h.c.}, \\
\label{Minus}
 &\lbrack c_{\k\sigma},c^\dag_{\k'\sigma'}\rbrack=-\eta_{\sigma\sigma'}\delta(\k-\k'),  \\
&    \lbrack c_{\k\sigma},c_{\k'\sigma'}\rbrack=0,
\end{align}
\label{Amu}%
\end{subequations} 
where $(k^\mu)=(\om{k},\k)$, $\om{k}=|\k|$, and
$\text{h.c.}$ stands for the Hermitian
conjugation \cite{remarkShell}. Moreover,  the  polarization $4$-vectors
satisfy  completeness
 and orthonormality relations
\begin{align}
&\sum_{\sigma=0}^3\eta_{\sigma\sigma}\epsilon^\mu(\k,\sigma)\epsilon^\nu(\k,\sigma)=\eta^{\mu\nu},\\
&\epsilon(\k,\sigma)\cdot\epsilon(\k,\sigma')=\eta_{\sigma\sigma'}.
\end{align}
They are chosen such that for transverse polarizations ($\sigma,\sigma'=1,2$)
\begin{align}
\label{pol1}
&\epsilon(\k,\sigma)=(0,\beps(\k,\sigma)), \\ 
&\beps(\k,\sigma)\cdot\beps(\k,\sigma')=\delta_{\sigma\sigma'}, \\ 
&\beps(\k,\sigma)\cdot\k=0,
\end{align}
while for  scalar ($\sigma=0$) and longitudinal  ($\sigma=3$) ones
\begin{align}
&\epsilon(\k,0)=(1,\0),\\
&\epsilon(\k,3)=(0,\k/\om{k}).
\label{pol2}
\end{align}

We also mention that 
Fock states,
obtained via  the repeated action of creation operators  $c_{\k\sigma}^\dag$ on the
vacuum state $|0\ra$, belong to the  indefinite-metric space, say ${\cal
H}$ from now on. It is so because
the prefactor on the right-hand side of (\ref{Minus}) is
negative for  $\sigma=\sigma'=0$. 

The Lorenz gauge constraint, amounting classically to $\partial\cdot A_\cl=0$, now takes the form 
\be
\langle\psi_\GB|\partial\cdot A|\psi'_\GB\rangle=0.
\label{GBzero}
\ee
This  holds when 
\be
L_{\k}\pGB = L_{\k}|\psi'_\GB\rangle = 0, \ L_{\k}=c_{\k3}-c_{\k0},
\label{GB}
\ee
which is supposed to happen for all $3$-momenta $\k$ \cite{remarkAll}. 
Such a  condition   follows from
\be
\partial\cdot A(z)= \ii\int
\frac{\d{k}}{(2\pi)^{3/2}}\sqrt{\frac{\om{k}}{2}} L_{\k} e^{-\ii k\cdot z} +\text{h.c.}
\label{dA}
\ee

States that are annihilated by operators $L_{\boldsymbol{k}}$
form the space  $\HGB$, where the  subscript  $\GB$ 
refers to Gupta and Bleuler. 
The elements of ${\cal H}_{GB}$ can be written as linear combinations of the states
\be
\begin{aligned}
\pGB= &|\psi_T\ra + \int \d{k} f(\k) L^\dag_{\k}|\psi_T\ra  \\
+&\int \d{k} \d{k'} f(\k,\k') L^\dag_{\k}  L^\dag_{\k'}|\psi_T\ra + \cdots, 
\end{aligned}
\label{LGB}
\ee
where $f(\k)$, $f(\k,\k')$, etc. are some functions
and $|\psi_T\ra$ is either the 
vacuum state or the state  containing  physical (transverse) photons only (Sec. 7.4 of \cite{Greiner}).
The  second, third, and the following terms on the right-hand side of
(\ref{LGB})
have zero norm, which can be seen by combining (\ref{GB}) with 
\be
[L_{\k},L^\dag_{\k'}]=0.
\label{LLzero}
\ee
Those states involve   unphysical (scalar and longitudinal) photons.
Their collection spans the space  $\HZERO$.

The physical states of the theory can be defined  by elements
of either   $\HGB$ or   $\HGB/\HZERO$. The former 
definition, e.g.,  can be found in Sec. 18 of 
\cite{Nakanishi1972} and Sec. I.5.3 of \cite{Haag}. The latter 
is employed  in   \cite{WightmanJMP1974}, discussing
the concept of equivalence between  those  vectors  from $\HGB$, which
differ  by a zero-norm vector
($\pGB$ and $|\psi_T\ra$ from 
(\ref{LGB}) belong to the same equivalence class for any choice of 
functions $f$). 
We adopt the former definition, calling
all states from $\HGB$ physical.
We will see implementation of this terminology right
below, where one more definition will be introduced.

The hermitian operator $O$ will be labeled  as  physical when 
$O\pGB\in\HGB$ for
all $\pGB\in\HGB$
\cite{WightmanJMP1974}. 
This requirement amounts to  saying that 
\be
[L_{\k},O]\pGB=0.
\label{OGB}
\ee

Having completed discussion of the covariant quantization of the electromagnetic field,
we are ready for writing down its   total canonical angular momentum operator 
\begin{subequations}
\begin{align}
&\J=\J_\spp+\J_\orbb+\J_\xi,\\
\label{JspQ}
&J^i_\spp= \int \d{z}  \varepsilon^{imn} F_{m0}A_n,\\
\label{JorbQ}
&J^i_\orbb= \int \d{z}  \varepsilon^{imn} z^m F_{j0}\partial_n A_j,\\
\label{Jxi}
&J^i_\xi=\int\d{z}\varepsilon^{imn}  z^m\partial\cdot A\partial_n
A_0,
\end{align}
\label{Jq}%
\end{subequations}
which can be obtained by combining  (\ref{Js}) with (\ref{Lcov}).
The following comments are pertinent to these expressions.

First, the time argument of $\J_\spp$, $\J_\orbb$, and $\J_\xi$ is
suppressed, which should not lead to any confusion as long as one remembers 
about remark \cite{remarkEqtimes}. We will proceed similarly with angular 
momentum operators 
discussed in Sec. \ref{QuantumII_sec}.

Second, the order of operators on the right-hand sides of (\ref{JspQ})--(\ref{Jxi})  does not matter, which
can be shown with (\ref{Amu}).

Third, we call $\J_\xi$ the  gauge-fixing angular momentum operator. The subscript
$\xi$ is used as  it is customarily associated with
covariant gauges. Note that such a type of angular momentum
 is absent in classical 
studies. 

Fourth, we will refer to $\J_\spp$, $\J_\orbb$, and $\J_\xi$ as angular
momentum operators (the same nomenclature will be applied 
to operators analyzed in Sec.
\ref{QuantumII_sec}). This is a slight abuse of terminology as none of them 
possesses all properties conventionally expected of  a genuine angular
momentum operator (e.g. none of them generates rotations of quantum 
fields). 
This will be discussed in detail in the following three subsections, 
where we will study various properties of these 
operators, paying special attention to the 
role of $\J_\xi$  in making $\J$ a genuine angular momentum
operator.

\subsection{Commutation relations}
\label{QuantumI_secA}
The first thing we will discuss  is  whether  gauge-fixing angular
momentum 
is needed for ensuring proper commutation relations of the total canonical 
angular momentum
operator
\be
[J^i,J^j]=\ii\varepsilon^{ijk} J^k.
\label{JJtot}
\ee

This can be investigated with the help  of  (\ref{Acan}),  leading
to  expressions that have to be integrated  by parts.
They will be handled   akin to the following integrals
\be
\begin{aligned}
\int\d{x}\d{y}\varepsilon^{imn}  x^m f(\x) g(\y)\frac{\partial}{\partial
x^n}\delta(\x-\y)& \\ 
\to-\int\d{x}  \varepsilon^{imn}x^m (\partial_nf) g&
\end{aligned}
\label{ip1}
\ee
and 
\be
\begin{aligned}
\int\d{x}\d{y}   f(\x) g(\y)\frac{\partial}{\partial
x^n}\delta(\x-\y)& \\ \to
\int\d{x}  [f\partial_ng-(\partial_nf) g]/2,&
\end{aligned}
\label{ip2}
\ee
where $f$ and $g$ are some  operators. 
It is a simple exercise to argue that  no boundary terms (surface integrals) are left out in (\ref{ip1}) and 
(\ref{ip2}).

In the end, we  find the following commutators  \cite{remarkEqtimes}
\be
[J^i_\chi,J^j_{\chi'}]=\ii\varepsilon^{ijk} J^k_\chi \delta_{\chi\chi'} \for 
\chi,\chi'=\spp,\orbb,\xi,
\label{Ccan}
\ee
where $\delta_{\chi\chi'}$ equals one for $\chi=\chi'$ and zero otherwise.
This shows that $\J_\spp$, $\J_\orbb$, and $\J_\xi$  satisfy  commutation relations expected 
of  angular momentum operators. Therefore,
with or without $\J_\xi$, (\ref{JJtot}) is satisfied.

Finally, it is worth mentioning that  (\ref{Ccan}) should not be taken for
granted. This is  best illustrated by the fact that 
Coulomb gauge versions of operators $\J_\spp$ and $\J_\orbb$ do not satisfy
such  commutation relations, which came as a surprise nearly three decades
ago \cite{EnkNienhuis1994}. We also mention that this well-known 
van Enk and Nienhuis result is 
discussed in review \cite{WakamatsuIntJModA2014}, where it
is stated  that  spin and orbital canonical angular
momentum operators of massless photons 
cannot  satisfy  proper angular momentum-type  commutation relations.
As (\ref{Ccan}) shows, such a conclusion does not apply to the 
covariantly-quantized  theory that we discuss.

\subsection{Generation of rotations}
\label{QuantumI_secB}
We will  take a closer look now at how various angular momentum operators 
 contribute to generation of rotations of 
tensor, vector, and scalar fields relevant to our studies. So, we will
investigate commutators 
\be
[V,J^i_\chi],
\ee
where $\chi$ is specified in  (\ref{Ccan}), 
$V=\partial_\mu A_\nu$, $A_\mu$, $\partial\cdot A$,  
and \cite{remarkEqtimes} is assumed.

We start by taking $V=A_\mu$. On the one hand, one can  show from the
Lorentz-transformation properties of $A_\mu$ that \cite{Greiner}
\be
\label{komVmu}
\ii[A_\mu(z),J^i]=(\z\times\bnabla)^iA_\mu-\varepsilon^{imn}\eta_{m\mu} A_n.§
\ee
On the other hand, using  (\ref{Acan}), we  can decompose 
(\ref{komVmu}) into 
\begin{align}
\label{V1}
&\ii[A_\mu(z),J^i_\spp]=-\varepsilon^{imn} \eta_{m\mu} A_n,\\
\label{V2}
&\ii[A_\mu(z),J^i_\orbb]=(\z\times\bnabla)^i (A_\mu - \eta_{0\mu} A_0),\\
&\ii[A_\mu(z),J^i_\xi]=\eta_{0\mu} (\z\times\bnabla)^i A_0.
\label{V3}
\end{align}
We see from these formulae that without the gauge-fixing contribution,
the total angular momentum operator does not  generate  rotations of $A_0$. It is so  because
 $A_0$ would commute with $\J$ if $\J$  would be stripped off  $\J_\xi$. 
This observation seems to be  intuitively plausible, if we take into account 
the fact that $A_0$ cannot be
canonically quantized  without the gauge-fixing modification of the Lagrangian
density. 

Building on these insights, it is tempting to 
speculate that  $\J_\xi$ is only needed  when rotations of fields
involving $A_0$ are considered. Such a speculation, however, is incorrect, which we
illustrate by choosing   $V=\partial_\mu A_\nu$.

So, we are dealing now with  
\be
\begin{aligned}
\label{komVmunu}
\ii[\partial_\mu A_\nu(z),&J^i]=
(\z\times\bnabla)^i\partial_\mu A_\nu\\
+&\varepsilon^{imn}(
\eta_{n\nu}\partial_\mu A_m - \eta_{m\mu} \partial_n A_\nu),
\end{aligned}
\ee
which can be derived in the same way as (\ref{komVmu}). We decompose it as
\be
\begin{aligned}
\ii[\partial_\mu A_\nu(z),&J^i_\spp]=\varepsilon^{imn} \eta_{n \nu}
\partial_\mu A_m\\
-& \eta_{0\mu}\eta_{0\nu} (\bnabla\times\A)^i\\
-&\varepsilon^{imn} \eta_{0\mu}\eta_{n \nu} \partial_m A_0,
\end{aligned}
\label{VV1}
\ee
\be
\begin{aligned}
\ii[\partial_\mu A_\nu(z),&J^i_\orbb]=
(\z\times\bnabla)^i\left[\partial_\mu A_\nu\right.\\
+&\eta_{0\mu}\eta_{0\nu}(\partial_0 A_0-\text{div}\A)\\
-&\left.  \eta_{0\mu}\partial_\nu A_0-\eta_{0\nu}\partial_\mu A_0\right]\\
-&\varepsilon^{imn}\eta_{m\mu}(\partial_n A_\nu - \eta_{0\nu}\partial_n A_0)\\
+&\eta_{0\mu}\eta_{0\nu}(\bnabla\times\A)^i,
\end{aligned}
\label{VV2}
\ee
\be
\begin{aligned}
\ii[\partial_\mu A_\nu(z),&J^i_\xi]=
(\z\times\bnabla)^i\left[\eta_{0\mu}\partial_\nu A_0\right.\\
+&\left.\eta_{0\nu}\partial_\mu A_0
+\eta_{0\mu}\eta_{0\nu}(\text{div}\A-\partial_0 A_0)\right]\\
+&\varepsilon^{imn}(\eta_{0\mu}\eta_{n\nu}\partial_m A_0 - \eta_{m\mu}\eta_{0\nu}\partial_n A_0).
\end{aligned}
\label{VV3}
\ee
Two remarks are in order now.

First, we see from these expressions that the gauge-fixing contribution is necessary
for reproducing (\ref{komVmunu}). Moreover, since (\ref{VV3}) is non-zero not
only for $\nu=0$ but also for
$\nu=1,2,3$, the importance of $\J_\xi$, in the context of rotations of field 
operators, cannot be restricted to expressions involving  $A_0$ only.

Second, we  easily  get from  (\ref{VV3}) that
\be
[F_{\mu\nu},J^i_\xi]=0.
\label{FFFJ}
\ee
This shows that  manifestly
gauge invariant operators, due to their sole  dependence on the electromagnetic tensor
$F_{\mu\nu}$, commute with $\J_\xi$.

Finally, regarding rotations of  scalar fields, we choose
$V=\partial\cdot A$
expecting on general grounds that 
\be
\ii[\partial\cdot A(z),J^i]=(\z\times\bnabla)^i\partial\cdot A,
\label{Sgen}
\ee
which we will use in Sec. \ref{QuantumI_secC}.
Straightforward calculations based on  (\ref{VV1})--(\ref{VV3}) 
lead to
\begin{align}
\label{VVV12}
&[\partial\cdot A,J^i_\spp]=[\partial\cdot A,J^i_\orbb]=0,\\
&\ii[\partial\cdot A(z),J^i_\xi]=(\z\times\bnabla)^i\partial\cdot A
\label{VVV3}
\end{align}
being in agreement with (\ref{Sgen}) solely thanks to $\J_\xi$'s presence in
$\J$.
We mention in passing that vanishing of the  commutator with $\J_\orbb$ should not
be generalized to other scalars. For example, it follows 
from (\ref{V2}) that $[A_\mu A^\mu,J^i_\orbb]\neq0$.

\subsection{Physical operator property}
\label{QuantumI_secC}
The objective of this section is to discuss whether  operators 
(\ref{JspQ})--(\ref{Jxi}), or some of their combinations, 
can be classified as  physical operators according to criterion 
(\ref{OGB}).

The first step, however, is to make sure  that the total 
canonical angular momentum operator
satisfies such a criterion. This   can be quickly verified with the help of 
\begin{subequations}
\begin{align}
&[L_{\k},\J]=-\ii(\k\times\bnabla_{\k}) L_{\k},\\
&(\bnabla_{\k})^i=\partial/\partial k^i,
\end{align}
\label{LJ}%
\end{subequations}
following  from (\ref{dA}) and (\ref{Sgen}).  
Combining   (\ref{GB}) and  (\ref{LJ}), we see
that $\J$  
is   a physical operator. This observation  agrees
with common sense understanding of what the 
total angular momentum operator 
should be.

The questions now are the following. Can  angular
momentum operators  $\J_\spp$, $\J_\orbb$, and $\J_\xi$ also be regarded  as
physical operators? 
If  not, can $\J_\spp + \alpha\J_\xi$, and so also 
$\J_\orbb+(1-\alpha)\J_\xi$, be proved to be   physical operators for some
 real $\alpha$?
Moreover,  can 
$\J_\spp+\J_\orbb$, superficially 
representing  total canonical angular momentum, 
be shown to be  a physical operator?

To proceed, we note that by using (\ref{Amu}) and 
properties of polarization vectors (\ref{pol1})--(\ref{pol2}), we get  
\be
\begin{aligned}
\label{LJxi}
&[L_{\k},\J_\xi]\pGB=\frac{\ii}{2} e^{2\ii\om{k}z^0}(\k\times\bnabla_{\k}) L^\dag_{-\k}\pGB,
\end{aligned}
\ee
\be
\begin{aligned}
[L_{\k},&\J_\spp]\pGB=\frac{\ii}{2\om{k}}
\sum_{\sigma=1}^2\big[\beps(\k,\sigma)\times\k\, c_{\k\sigma}\\
&+e^{2\ii\om{k}z^0}\k\times\beps(-\k,\sigma) c^\dag_{-\k\sigma}\big]\pGB,
\end{aligned}
\ee
\be
\begin{aligned}
&[L_{\k},\J_\orbb]\pGB=-
[L_{\k},\J_\spp+\J_\xi]\pGB,
\end{aligned}
\label{kkllkkll}
\ee
where  
(\ref{kkllkkll})  has been obtained  without using the
above-established fact that 
\be
[L_{\k},\J]\pGB=0
\label{LJzz}
\ee
and $|\psi_\GB\ra$ satisfies   (\ref{GB}).
Several  comments are in order now.

First, $z^0$-dependence comes from the fact that  the electromagnetic vector 
potential, entering
definitions of angular momentum operators (\ref{JspQ})--(\ref{Jxi}),
depends on $z=(z^0,\z)$.

Second,  none of $\J_\spp$, $\J_\orbb$, and $\J_\xi$  is  a physical 
operator
\be
[L_{\k},\J_\chi]\pGB\neq0 \for \chi=\spp,\orbb,\xi.
\label{LJchi}
\ee
Moreover, it is easy to see that no $\alpha$ leads to vanishing of
$[L_{\k},\J_\spp + \alpha\J_\xi]\pGB=-[L_{\k},\J_\orbb +  (1-\alpha)\J_\xi]\pGB$.

Third, just because (\ref{LJxi}) is non-zero,  the gauge-fixing contribution is 
necessary for making $\J$ a physical operator.
This means that $\J_\spp+\J_\orbb$ is not   a physical  operator in 
the covariantly-quantized 
theory. 
All in all,  answers to  the above-mentioned  questions are  unfortunately negative. 

It is perhaps worth to mention that the fact that operators 
$\J_\spp$, $\J_\orbb$, and $\J_\xi$  are  gauge-variant
does not imply  (\ref{LJchi}). In fact, the sum of all of them,
the total canonical angular momentum operator $\J$, is also gauge-variant, but it satisfies 
(\ref{LJzz}).
We mention in passing that gauge non-invariance of  
$\J$  should not be surprising given the fact that 
we work with the gauge-fixed theory (see also \cite{LeaderPRD2011}).

Finally, we note that  
\be
[L_{\k},F_{\mu\nu}]=0.
\label{LF}
\ee
Therefore,   manifestly 
gauge invariant operators, in the sense mentioned below (\ref{FFFJ}), 
can be labeled as physical.

\section{Quantum considerations: Belinfante-Rosenfeld  expressions}
\label{QuantumII_sec}
We find by combining 
 (\ref{emtensor}), (\ref{chi3}), (\ref{tildeJ}), and (\ref{Lcov}) that 
the covariant-gauge quantum version of  Belinfante-Rosenfeld expression (\ref{Jinv}) is
\begin{subequations}
\begin{align}
&\tJ=\J_\fl  + \J_\DIV + \J_\xi,\\
\label{ttmm1}
&J^i_\fl=\int\d{z} \varepsilon^{imn} z^m F_{0j}F_{jn},\\
&J^i_\DIV=\int\d{z} \varepsilon^{imn}z^m A_n\partial_jF_{0j}.
\label{Jidiv}
\end{align}
\label{Jtot2}%
\end{subequations}
Several remarks are in order now.

First, $\J_\fl$ is called field angular momentum to underscore its 
dependence on the electromagnetic field. It is given by the right-hand side of 
(\ref{Jinv}) stripped off the symbol $\cl$.
It is of special interest because it is  gauge invariant 
unlike $\J_\spp$ and $\J_\orbb$.
 Moreover, there is interesting classical 
 physics associated with it (see
the discussion of  the  Feynman's disk paradox in Secs. 17-4 and 27-6 of \cite{FeynmanVolII} and
\cite{LombardiAmJPhys1983,BahderAmJPhys1985,MaAmJPhys1986}).

Second, $\J_\DIV$, which is  absent in classical 
considerations (Sec. \ref{Classical_sec}), appears here  because 
the covariantly-quantized electric  field operator
satisfies the modified  Gauss' law 
\be
\text{div}\E=-\partial_0(\partial\cdot A),
\label{Gcov}
\ee
where the  right-hand side of (\ref{Gcov}) is non-zero 
due to (\ref{dA}). 
Non-vanishing of $\J_\DIV$, which is actually  counterintuitive  in the absence
of charged particles,
may be also seen as a by-product of the fact that 
$\partial\cdot A=0$ cannot be imposed on the operator level. 
We mention in passing that the symbol $\J_\DIV$
refers to the fact that there is the divergence operator in (\ref{Jidiv}).

Third, densities of the canonical and Belinfante-Rosenfeld total angular 
momentum operators differ by the $3$-divergence
\be
\J-\tJ= \int \d{z} \partial_j\BB{E^j(\z\times\A)}.
\label{diffJJ}
\ee
 As will be seen below, the properties 
of $\tJ$, which we investigate, are the same as those of $\J$.
Therefore,  there is no need  for discussion of 
such a boundary term  in the context of our studies.

Finally, we mention that 
the order of operators on the right-hand sides of
(\ref{ttmm1}) and (\ref{Jidiv}) does not matter, which can be 
verified with (\ref{Amu}).

The following  discussion will be  organized similarly  as in Secs.
\ref{QuantumI_secA}--\ref{QuantumI_secC}, which will allow us for keeping  it concise.

\subsection{Commutation relations}
\label{CommBel_sec}
We find that  commutators of 
angular momentum operators from  (\ref{Jtot2})
are given by  \cite{remarkEqtimes}
\be
\begin{aligned}
[J^i_\fl,&J^j_\fl]=\ii\varepsilon^{ijk}J^k_\fl\\
&+\ii
\int\d{z}\varepsilon^{imn}\varepsilon^{jks}z^m z^k F_{sn}\text{div}\E,
\end{aligned}
\label{div1}
\ee
\be
\begin{aligned}
[J^i_\fl,&J^j_\DIV]=-\ii\int\d{z}\varepsilon^{imn}\varepsilon^{jks}z^m z^k F_{sn}\text{div}\E,
\end{aligned}
\label{div2}
\ee
\be
[J^i_\fl,J^j_\xi]=0,
\ee
\be
\begin{aligned}
[J^i_\DIV,&J^j_\DIV]=\ii\varepsilon^{ijk}J^k_\DIV\\
&+\ii
\int\d{z}\varepsilon^{imn}\varepsilon^{jks}z^m z^k F_{sn}\text{div}\E,\\
\end{aligned}
\label{div3}
\ee
\be
[J^i_\DIV,J^j_\xi]=0,
\label{divLast}
\ee
where $[J^i_\xi,J^j_\xi]$ is not listed as it is given in  (\ref{Ccan}).
The following  observations are pertinent to these expressions. 

First, there are no ordering ambiguities 
on the right-hand sides of
(\ref{div1}), (\ref{div2}), and
(\ref{div3}), which can be shown with (\ref{Amu}).

Second, we see from these expressions that neither $\J_\fl$ nor $\J_\DIV$
satisfies  algebra
expected of angular momentum operators as
\be
[J^i_\chi,J^j_{\chi'}]\neq\ii\varepsilon^{ijk}J^k_\chi\delta_{\chi\chi'} \for 
\chi,\chi'=\fl,\DIV
\ee
 due to $\text{div}\E\neq0$ (\ref{Gcov}).
This   differs from what we have found in Sec. \ref{QuantumI_secA}, where
every component of the
 total angular momentum operator
  individually satisfied proper commutation relations.

Third, the proper commutation relations are satisfied by 
 $\J_\fl  + \J_\DIV$ and  $\J_\fl +  \J_\DIV + \J_\xi$.
So, the total  Belinfante-Rosenfeld angular momentum operator does not need
 the $\J_\xi$  contribution   for 
 having  proper   commutation relations.
 It does need, however, the  contribution of $\J_\DIV$, which is 
 a bit
surprising in the system without charged 
particles.

\subsection{Generation of rotations}
Proceeding similarly  as in Sec. \ref{QuantumI_secB},  we compute 
\be
\begin{aligned}
\ii[A_\mu(z),&J^i_\fl]= (\z\times\bnabla)^i (A_\mu - \eta_{0\mu}  A_0)\\
-&\varepsilon^{imn} z^m (\partial_\mu A_n-\eta_{0\mu} \partial_0 A_n),
\end{aligned}
\label{startt}
\ee

\be
\begin{aligned}
\ii[A_\mu(z),&J^i_\DIV]=-\varepsilon^{imn} \eta_{m\mu} A_n\\
+&\varepsilon^{imn} z^m (\partial_\mu A_n-\eta_{0\mu} \partial_0 A_n),
\end{aligned}
\ee
and 
\be
\begin{aligned}
\ii[\partial_\mu A_\nu(z),&J^i_\fl]=
\varepsilon^{imn}z^m\partial_\mu F_{n\nu}\\
-&\varepsilon^{imn}\eta_{m\mu}F_{n\nu}\\
+&\varepsilon^{imn}(\eta_{0\mu}\eta_{n\nu} F_{0m}-\eta_{m\mu}\eta_{0\nu}F_{0n})\\
+&\varepsilon^{imn}z^m(\eta_{0\mu}\partial_\nu F_{0n}+\eta_{0\nu}\partial_\mu F_{0n})\\
-&\varepsilon^{imn}z^m\eta_{0\mu}\eta_{0\nu}(\partial_0 F_{0n} + \partial_j F_{jn})\\
+&\varepsilon^{imn}z^m\eta_{0\mu}\eta_{n\nu}\partial_j F_{0j}\\
+&2\eta_{0\mu}\eta_{0\nu}(\bnabla\times\A)^i,
\end{aligned}
\ee
\be
\begin{aligned}
\ii[\partial_\mu A_\nu(z),&J^i_\DIV]=\varepsilon^{imn}z^m\partial_\mu\partial_\nu A_n\\
-&\varepsilon^{imn}(\eta_{m\mu}\partial_\nu A_n - \eta_{n\nu}\partial_\mu A_m)\\
+&\varepsilon^{imn}(\eta_{m\mu}\eta_{0\nu}\partial_0 A_n - \eta_{0\mu}\eta_{n\nu}\partial_0 A_m)\\
-&\varepsilon^{imn}z^m(\eta_{0\mu}\partial_\nu\partial_0 A_n + \eta_{0\nu}\partial_\mu\partial_0 A_n)\\
+&\varepsilon^{imn}z^m\eta_{0\mu}\eta_{0\nu}(\partial_0\partial_0 A_n+\partial_j\partial_j A_n)\\
-&\varepsilon^{imn}z^m\eta_{0\mu}\eta_{n\nu}\partial_j F_{0j}\\
-&2\eta_{0\mu}\eta_{0\nu}(\bnabla\times\A)^i.
\end{aligned}
\label{endd}
\ee

These two sets of 
results--when combined with (\ref{V3}) and (\ref{VV3})--can be used for showing that 
the total Belinfante-Rosenfeld angular momentum operator 
generates rotations of $A_\mu$, $\partial_\mu A_\nu$, and $\partial\cdot A$
(see Sec. \ref{QuantumI_secB} for the discussion of how this can be done).

One can also show with  (\ref{startt})--(\ref{endd})
that not only 
$\J_\DIV$'s 
but also 
$\J_\xi$'s
contribution to $\tJ$ is indispensable  for 
making it   a proper generator of rotations of $A_\mu$ and $\partial_\mu
A_\nu$. As far as rotations of $\partial\cdot A$ are concerned, we mention
that $[\partial\cdot A,J^i_\fl]=[\partial\cdot A,J^i_\DIV]=0$,
which is very much similar to the situation encountered in 
Sec. \ref{QuantumI_secB}.

All in all, it is worth  stressing that 
$\J_\fl$ does not generate rotations of above-discussed fields.

\subsection{Physical operator property}

With the help of (\ref{LF}), we immediately see that 
$\J_\fl$ is a physical operator, which agrees
with common sense expectations 
following from its  manifest    gauge invariance.

The same cannot be said about $\J_\DIV$, for which we get
\be
[L_{\k},\J_\DIV]\pGB=-\frac{\ii}{2} e^{2\ii\om{k}z^0}(\k\times\bnabla_{\k}) L^\dag_{-\k}\pGB,
\label{LJdiv}
\ee
where  $\pGB$ satisfies   (\ref{GB}).
Combining it with  (\ref{LJxi}) and (\ref{LF}), we find that 
 total Belinfante-Rosenfeld
angular momentum operator (\ref{Jtot2})  is a physical operator
due to  
\be
[L_{\k},\tJ]\pGB=0.
\ee
This is an expected result. What is not so obvious, however,
is that it would not be the case  if just  one of 
the gauge-fixing-related angular momentum operators, 
$\J_\xi$ or $\J_\DIV$, would be missing in $\tJ$.

\section{Physical matrix elements of gauge-fixing-related operators}
\label{Gauge_sec}
Terms in above-discussed  operators, which originate from the
gauge-fixing modification of the Lagrangian density, 
have one common feature. Namely, they involve the 
operator $\partial\cdot A$. This is easily seen in $\J_\xi$  and it
can be also  noticed in $\J_\DIV$  after employment of  (\ref{Gcov}).
If we now note that matrix elements of such an operator vanish in 
physical states of the theory (\ref{GBzero}), 
then one may wonder whether the very same   result is obtained after replacement of $\partial\cdot A$
in (\ref{GBzero}) by a ``composite'' operator involving $\partial\cdot A$. 
If that would be the case, then  physical matrix elements of either 
 gauge-fixing-originated operators or   their products with other operators
could be set to zero.

The above speculation is incorrect, i.e. 
physical matrix elements of operators involving  $\partial\cdot A$
are not necessarily zero, and we find it instructive to explain why
it is so. To begin, however, we will  play {\it advocatus diaboli} to get to the
bottom of the issue. 
We consider
\be
\la\psi_\GB|O \partial\cdot A|\psi_\GB'\ra,
\label{L1}
\ee
where $O$ is some operator, and insert 
between $O$ and $\partial\cdot A$ 
the following resolution of identity in ${\cal H}$
\begin{subequations}
\begin{align}
\label{not1}
&\mathbbm{1}\dot{=}\sum_{|\psi_\GB''\ra\in\HGB} |\psi_\GB''\rl\psi_\GB''| + 
\sum_{|\phi\ra\notin\HGB} |\phi\rl\phi|,\\
&
\la\psi_\GB''|\phi\ra\dot{=}0 \for |\psi_\GB''\ra\in\HGB, |\phi\ra\notin\HGB,
\label{not2}
\end{align}
\label{not12}%
\end{subequations}
where the dot over the  equality symbol  indicates that such an 
expression will be disputed below. 
Leaving doubts aside, we arrive at
\begin{subequations}
\begin{align}
\la\psi_\GB|&O \partial\cdot A|\psi_\GB'\ra\dot{=}\nonumber\\
\label{z1}
&\sum_{|\psi_\GB''\ra\in\HGB}\la\psi_\GB|O|\psi_\GB''\rl\psi_\GB''| \partial\cdot A|\psi_\GB'\ra\\
\label{z2}
+&\sum_{|\phi\ra\notin\HGB}\la\psi_\GB|O|\phi\rl\phi| \partial\cdot A|\psi_\GB'\ra\\
\dot{=}&0.
\label{z3}
\end{align}
\label{z}%
\end{subequations}
The result reported in (\ref{z3}) follows from the following
observations. First,  (\ref{z1}) vanishes because of  (\ref{GBzero}). Second, due
to (\ref{dA})--(\ref{LLzero}), or simply $[L_{\k},\partial\cdot A]=0$, 
we have  $\partial\cdot A|\psi_\GB'\ra\in\HGB$, 
and then (\ref{z2}) vanishes as a consequence of  (\ref{not2}).

The most elementary counterexample to the above-obtained result 
is the following \cite{remarkStrocchi,remarkShell}
\be
\begin{aligned}
\la0|A_\mu(x) & \partial\cdot A(y)|0\ra=\\
&-\frac{\ii}{2}\int \ddd{k} \frac{k_\mu}{\om{k}}e^{-\ii k\cdot(x-y)}\neq0.
\end{aligned}
\label{nzero}
\ee
The disagreement between (\ref{z}) and (\ref{nzero}) is the result of 
the tacit assumption that  the resolution
of identity  in the  indefinite-metric space has the same structure as in a
Hilbert space. This is actually not what is  happening {\it if}
one  insists
on the use of explicit projectors onto  states from $\HGB$, 
 which we discuss below.

Introducing some notation, we write
\be
\mathbbm{1}=\mathbbm{1}_{0\otimes0}+\mathbbm{1}_{1\otimes0}+\mathbbm{1}_{0\otimes1}+
\mathbbm{1}_{2\otimes0}+\mathbbm{1}_{1\otimes1}+\mathbbm{1}_{0\otimes2}+
\dots,
\label{111}
\ee
where $\mathbbm{1}_{n\otimes m}$ denotes the unit operator in the subspace with
the number of physical (unphysical) photons equal to $n$ ($m$).

In the zero-photon sector,  one trivially has
\be
\mathbbm{1}_{0\otimes0}=|0\rl0|.
\label{00}
\ee

In the one-photon sector, the basis states are
\be
|\k\sigma\ra=c^\dag_{\k\sigma}|0\ra,
\ee
where the polarization index
$\sigma=0,1,2,3$. 
Taking into account the fact that
$\la\k\sigma|\k'\sigma'\ra=-\eta_{\sigma\sigma'}\delta(\k-\k')$, 
one has
\begin{align}
\label{10}
&\mathbbm{1}_{1\otimes0}=\int\d{k}\sum_{\sigma=1}^2|\k\sigma\rl\k\sigma|,\\
\label{01}
&\mathbbm{1}_{0\otimes1}=\int\d{k} (|\k3\rl\k3| - |\k0\rl\k0|),
\end{align}
where the negative sign in front of the scalar-photon projector
reminds us that we work in the indefinite-metric space.

To arrive at the expression that is more relevant in the context of our discussion of (\ref{not12})--i.e.
the one that has explicit projectors on {\it all} states 
from $\HGB$--we have rewritten (\ref{01}) in the following way 
\be
\begin{aligned}
\mathbbm{1}_{0\otimes1}=\int\d{k}&\big(
a_\alpha|\Phi_1(\k)\rl\Phi_1(\k)|\\
+&|\Phi_1(\k)\rl\Rzymskie{1}_\alpha(\k)|
+|\Rzymskie{1}_\alpha(\k)\rl\Phi_1(\k)|
\big),
\end{aligned}
\label{01new}
\ee
where  $\alpha\in\mathbb{C}\setminus\{-1\}$ is assumed and 
\begin{align}
\label{PHI1}
&|\Phi_1(\k)\ra=L^{\dag}_{\k}|0\ra=|\k3\rangle-|\k0\rangle,\\
&|\Rzymskie{1}_\alpha(\k)\ra=\frac{|\k3\ra}{1+\alpha}+\frac{\alpha|\k0\ra}{1+\alpha}, \\
&a_\alpha=\frac{|\alpha|^2-1}{|\alpha+1|^2}.
\label{aalfa}
\end{align}
The main difference between (\ref{01}) and (\ref{01new}) is that the latter 
singles out  (\ref{PHI1}), the 
``only'' zero-norm state in  the one-photon subspace of $\HGB$
\cite{remarkOnly}. Moreover, we mention  that for real $\alpha\neq-1$,
the expression for 
$a_\alpha$ reduces to $(\alpha-1)/(\alpha+1)$,
which is a special case of the M\"obius transformation. This simple observation 
suggests that there might be some geometric interpretation of
(\ref{01new}).

Next, we note that (i) the  one-photon sector of $\HGB$
is spanned by mutually orthogonal  states $|\k1\ra$, $|\k2\ra$, and $|\Phi_1(\k)\rangle$; (ii)  
 $|\Rzymskie{1}_\alpha(\k)\ra\notin\HGB$;
(iii) $\la\Phi_1(\k)|\Rzymskie{1}_\alpha(\p)\ra=\delta(\k-\p)\neq0$;  
(iv)
states 
\be
|\k1\ra, |\k2\ra,  |\Phi_1(\k)\rangle, |\Rzymskie{1}_\alpha(\k)\ra
\label{basbas}
\ee
form a basis in the $\k$-momentum sector
of the one-photon subspace of
the indefinite-metric  space \cite{remarkAll}.

Combining these  observations, 
one can easily see fundamental differences between
(\ref{not12}) and the correct resolution of identity in ${\cal H}$.
Namely, neither (\ref{not1}) matches the sum of  (\ref{10}) and (\ref{01new})
nor  orthogonality condition (\ref{not2}) is satisfied by all basis states
(\ref{basbas}).
These unusual characteristics, absent in a Hilbert space, 
are not limited to the one-photon sector of the theory. We  illustrate this 
remark in 
 Appendix \ref{Two_app}, extending the above considerations to the two-photon sector. 
One can also find there explanation of the steps leading to derivation of (\ref{01new}) and 
 further insights into  decomposition of unity in the
indefinite-metric space of our  system.

We can now apply these formulae to computation of (\ref{nzero}).
Inserting (\ref{111}) between $A_\mu$ and
$\partial\cdot A$, we instantly  arrive with the help of (\ref{GBzero})
at 
\be
\begin{aligned}
\la0|&A_\mu(x) \partial\cdot A(y)|0\ra=\\
&\int \d{k}\la0|A_\mu(x)|\Phi_1(\k)\ra
\la\Rzymskie{1}_{\alpha}(\k)|\partial\cdot A(y)|0\ra,
\end{aligned}
\ee
which 
reproduces (\ref{nzero}) after straightforward algebraic manipulations.

If we now turn our attention to \cite{LeaderPRD2011}, 
we will find there 
the statement  that gauge-fixing-originated contributions to operators, 
representing  densities of  canonical and 
Belinfante-Rosenfeld energy-momentum tensors,
have vanishing physical matrix elements in covariantly-quantized 
electrodynamics (cQED).
To argue that it is so, insertion of a complete set of physical states
between field operators is mentioned without an explanation of what it
actually amounts to. Such a procedure of proving the above statement 
cannot be  correct, which can be  shown 
with the results that we have already presented 
in this section.
We briefly comment on that below,  discussing 
the simplest counterexample we  can think of.

The free-field Feynman gauge versions of the operators discussed in  \cite{LeaderPRD2011}
are \cite{remarkB}
\be
t^{\mu\nu}_\text{can}(\text{Gf})=
-\partial\cdot A\partial^\nu A^\mu+\frac{\eta^{\mu\nu}}{2}(\partial\cdot A)^2
\label{Candelta}
\ee
and 
\be
t^{\mu\nu}_\text{bel}(\text{Gf})=
\partial^\mu(\partial\cdot A) A^\nu+
\partial^\nu(\partial\cdot A) A^\mu
+\frac{\eta^{\mu\nu}}{2}(\partial\cdot A)^2.
\label{Beldelta}
\ee
Using (\ref{nzero}), we  find that \cite{remarkShell} 
\be
\begin{aligned}
\la0|t^{\mu\nu}_\text{bel}(\text{Gf})|0\ra
&=2\la0|t^{\mu\nu}_\text{can}(\text{Gf})|0\ra\\
&=\int \ddd{k}\frac{k^\mu k^\nu}{\om{k}} \neq0 \for \mu=\nu,
\end{aligned}
\label{0d0}
\ee
which  disagrees with (59)   from \cite{LeaderPRD2011}. Note 
that the fact
that \cite{LeaderPRD2011} deals with the  interacting theory, cQED, does not 
explain this disagreement.  

On the one hand,  (\ref{0d0})  illustrates the point we are trying to make 
that employment of 
the resolution of identity in the indefinite-metric space is far from 
trivial and it must have been misunderstood in some way in \cite{LeaderPRD2011}. 
We also mention that insertion of (\ref{not1}) between field operators in
(\ref{Candelta}) and (\ref{Beldelta}) 
sets (\ref{0d0}) to zero, reproducing the result reported in \cite{LeaderPRD2011}. 
The same wrong null result is obtained if
one inserts, during computation of (\ref{0d0}), only the first term of (\ref{not1}) 
between the field operators. 

On the other hand, (\ref{0d0})  suggests re-examination of 
conclusions first discussed by the end of Sec. IV of  \cite{LeaderPRD2011} 
and then  reiterated in Sec. 3.1.1 of \cite{LeaderPhysRep2014short}.

If we now turn our attention to angular momentum 
operators involving $\partial\cdot A$, we will 
find that 
\be
\la\psi_\GB|\J_\xi|\psi_\GB'\ra = \la\psi_\GB| \J_\DIV|\psi_\GB' \ra = 0.
\label{zxid}
\ee
 This is seen by using
(\ref{Amu})--(\ref{LLzero}) to show that 
\be
\begin{aligned}
&\la\psi_\GB|
\partial\cdot A  
\partial_n A_0
|\psi_\GB'\ra\\
&=-\la\psi_\GB| A_n\partial_0(\partial\cdot A)|\psi_\GB'\ra\\
&=\la\psi_T|\psi'_T\ra \int\ddd{k}\frac{k^n}{2}=0,
\end{aligned}
\label{xiZZ}
\ee
where  $|\psi_\GB'\ra$   and   $|\psi_T'\ra$   are related to each other via 
the ``primed'' version of (\ref{LGB}).

It should be now understood that (\ref{zxid}) is obtained under the  
assumption that the  free electromagnetic field
is considered.
In the interacting theory,  cQED, 
results such as (\ref{zxid}) give tree-level (zeroth-order) expressions
and perturbative (loop) calculations are needed for checking whether 
there are radiative corrections to them.
For example, a one-loop correction to gauge-fixing angular momentum, 
in the cQED ground state describing 
the electron at rest, depends on \cite{BDall}
\be
\begin{aligned}
&\la0|\T\N{(\J^I_\xi)^i}  A^I_\mu(x) A^I_\nu(y)|0\ra=\\
&\int\d{z} \varepsilon^{imn}  z^m 
\la0|\T\N{\partial\cdot A_I(z)\partial_n A^I_0(z)} A^I_\mu(x) A^I_\nu(y)|0\ra=\\
&\int\frac{\d{z}\dd{p}\dd{q}}{(2\pi)^8} \frac{\varepsilon^{imn}z^m p_\mu q_n\eta_{0\nu}}{(p^2+\izero)(q^2+\izero)}
e^{\ii p\cdot(z-x)+\ii q\cdot(z-y)}\\
&+\xymunu,
\end{aligned}
\label{Ang}
\ee
where $\T$ is the time-ordering operator,
$\N{ \ }$ 
denotes normal ordering, 
and $A^\mu_I$ is the interaction-picture vector potential, which can be written  just as (\ref{Amu}). 
The answer to the question,  whether one-loop radiative corrections 
make   the  expectation value of 
$\J_\xi$   non-zero, relies then  on 
fermionic contributions, which are contracted with (\ref{Ang}). 
 It 
is shown in \cite{BDall} that $\la\J_\xi\ra$ is non-zero 
in the cQED ground state describing the electron at rest.
It is also shown there that such a gauge-fixing contribution 
is needed in order to obtain 
the result 
$\la \J_\text{total}\ra= 
\boldsymbol{\hat s}/2+O(\alpha^2)$
for the electron's total angular momentum in such a state,
where $\J_\text{total}$ stands for the  total canonical 
angular momentum operator of cQED, 
the unit $3$-vector $\boldsymbol{\hat s}$ denotes polarization of the electron's spin,
$\alpha$ is the fine-structure constant, and 
 $O(\alpha^2)$
reflects the fact that two- and higher-loop
radiative corrections  were   
not studied  in \cite{BDall}.

\section{Summary}
\label{Summary_sec}
If one wants to study
the total angular momentum operator of the free  electromagnetic field,
then the traditional discussion of  angular momentum decompositions
makes the
impression that one has to choose between the canonical 
\be
\J_\spp+\J_\orbb,
\label{J111}
\ee
the Belinfante-Rosenfeld
\be
\J_\fl,
\label{J222}
\ee
or some other expressions  free from  explicit gauge-fixing-originated 
contributions (see e.g. review  \cite{WakamatsuIntJModA2014}).
The reality, however, is that in the covariantly-quantized theory 
neither  (\ref{J111}) nor  (\ref{J222})  
can  be considered as a total angular momentum
operator.

This has been discussed in Secs. \ref{QuantumI_sec} and \ref{QuantumII_sec},
where it has been  shown that neither (\ref{J111}) nor (\ref{J222}) 
generates rotations of quantum fields. It can be also found there that 
these expressions have some  complementary deficiencies. For example,
(\ref{J111}) has an unwelcome feature that it 
takes  physical states of the theory outside of the physical sector 
of the state space, whereas  (\ref{J222}) does not lead to such complications. 
At the same time, (\ref{J222}) does not satisfy 
commutation relations expected of 
a total  angular momentum operator while
(\ref{J111}) does.
These are basic reasons why a   genuine total angular momentum operator
of the  covariantly-quantized electromagnetic field, 
such as   (\ref{Jq}) or  (\ref{Jtot2}),
needs the  explicit gauge-fixing
contribution(s).

It should be also recognized that 
gauge-fixing 
is implicitly encoded in (\ref{J111}) and (\ref{J222}) once 
these operators are defined via 
the vector potential, which  satisfies  commutation
relations (\ref{Acan}) or
has
Fourier expansion (\ref{Amu}). It is so because both  (\ref{Acan}) and (\ref{Amu})
 owe their elegant  covariant form  
to the gauge-fixing modification of the
Lagrangian density.
Looking from this perspective, the need for inclusion of gauge-fixing
corrections to (\ref{J111}) and (\ref{J222}) 
can be seen as a consistency requirement.
Namely, one should not benefit from 
gauge-fixing when it comes to field operators and pretend that it is 
non-existent while investigating   total angular momentum operators.

The above discussion can be easily extended to quantum electrodynamics, where 
physical insights are typically gained via perturbative expansions as it is 
an interacting theory. Those 
are oftentimes carried out with the following interaction-picture  photon 
 propagator
\be
\la0|\T A^I_\mu(x) A^I_\nu(y)|0\ra= \int
\dddd{k}\frac{-\ii\eta_{\mu\nu}}{k^2+\izero}e^{-\ii k\cdot(x-y)},
\label{FIP}
\ee
having such an elegant covariant  form due to the gauge-fixing modification of the
Lagrangian density. Thus, employment of (\ref{FIP}) in actual calculations
automatically implies that the Lagrangian  is gauge fixed and that 
some observables derived  from it acquire 
gauge-fixing-originated contributions absent in the classical theory.

Our studies  should not suggest that only those angular momentum 
operators that satisfy all
criteria investigated in Secs. \ref{QuantumI_sec} and \ref{QuantumII_sec} are 
physically interesting. In fact,  there are different forms
of angular momentum. The   operators representing them cannot 
possibly have  the same properties as a total angular momentum operator,
if they are just a part of it. 
Still, they can be  in principle measurable and carry out useful
information about the studied system 
(see e.g. \cite{EnkNienhuis1994,WakamatsuIntJModA2014} for  a
similar viewpoint). For this to happen, they do not 
need to have angular momentum-like commutation relations and do not have to
properly generate rotations. We assume, however, that they should satisfy 
physical observable criterion. As far as this work is concerned, 
(\ref{J222}) is
an example of such an operator, which is additionally also gauge invariant 
(manifest gauge invariance of $\J_\fl$ actually implies its physical operator property).
We mention in passing that its studies in cQED can be found 
in \cite{BDfield}, see also \cite{remarkBDfield}.

Having mentioned gauge invariance, it should be also said that 
total angular momentum operators,  (\ref{Jq}) and  (\ref{Jtot2}), are  gauge variant. 
This  is not a controversial  feature 
because gauge-fixing explicitly breaks 
invariance of the theory with respect to arbitrary gauge transformations (see
also discussion in \cite{LeaderPRD2011}). Note that this does not mean
that matrix elements  of such operators, in physical states of the theory, 
are gauge variant. 
 For example, 
it was explicitly shown in \cite{BDall}  that the one-loop 
 expectation value of the total canonical angular momentum operator, 
in the  cQED  state describing the electron
at rest,   equals $1/2$ in {\it all} covariant gauges.

These remarks bring us to the discussion of physical matrix elements 
of the so-called 
gauge-fixing-related  operators, i.e., the ones involving $\partial\cdot A$ in the
context of our studies.
We have shown  how  one can reach incorrect  results 
for such matrix elements, if one does not take into account subtleties of the 
resolution of identity in the
indefinite-metric space of the covariantly-quantized electromagnetic field.
Namely,   one has to be careful  when 
one  wants to  use
explicit projectors onto the physical sector of such a space.
If this is the case, then    not-so-intuitive expressions for the unit operator,  in the one- and
two-photon subspaces,  take the form discussed   in Sec. \ref{Gauge_sec} and Appendix
\ref{Two_app}. 
Apart from technical insights, these considerations convey 
the message that  
it should not be {\it a priori} assumed that 
physical matrix elements of operators, having
gauge-fixing-related components,  vanish.
It would be interesting to investigate similar issues in quantum
chromodynamics.

\section*{ACKNOWLEDGMENTS}
I would like to dedicate this work to the memory of Andrzej Grzebie\'n, 
who passed away during the course of these studies.
This work has been  supported by the Polish National Science Centre (NCN)
grant 2019/35/B/ST2/00034. 

\appendix
\section{Two-photon sector of $\mathbbm{1}$}
\label{Two_app}
We discuss here the resolution of identity in the two-photon subspace of the
indefinite-metric space ${\cal H}$. On the one hand, 
this allows  for explicit illustration 
of the fact that  the abnormal  form        
of the unit operator, 
which we have encountered in Sec. \ref{Gauge_sec}, is not specific to the 
one-photon sector of ${\cal H}$.
On the other hand, the following studies clearly illustrate
the logic behind derivation of (\ref{01new}), which has not been explained in
Sec. \ref{Gauge_sec}.
Some general remarks about our studies of 
the resolution of identity  are
provided by the end of this Appendix.

The  two-physical-photon projector is given by 
\begin{subequations}
\begin{align}
&\mathbbm{1}_{2\otimes0}=\int\d{k}\d{k'}
\sum_{\sigma,\sigma'=1}^2
|\k\sigma\k'\sigma'\rl\k\sigma\k'\sigma'|,\\
&|\k\sigma\k'\sigma'\ra= \frac{1}{\sqrt{2}}c^\dag_{\k\sigma}c^\dag_{\k'\sigma'}|0\ra.
\label{qwsa1}
\end{align}
\label{20}%
\end{subequations}

The mixed physical-unphysical-photon projector reads 
\be
\mathbbm{1}_{1\otimes1}=\int\d{k'}\sum_{\sigma'=1}^2c^\dag_{\k'\sigma'}\mathbbm{1}_{0\otimes1}c_{\k'\sigma'}.
\label{11}
\ee

A more interesting situation is encountered in the  two-unphysical-photon
sector, where 
\be
\begin{aligned}
\mathbbm{1}_{0\otimes2}=\int  \d{k}\d{k'} 
\big(
&|\k3\k'3\rl\k3\k'3|
+|\k0\k'0\rl\k0\k'0|\\
-&|\k0\k'3\rl\k0\k'3|
-|\k3\k'0\rl\k3\k'0|
\big).
\end{aligned}
\label{1102}
\ee

In this sector of the indefinite-metric space, 
there is one state from $\HGB$, the zero-norm state  (\ref{P2}).
Our goal now is to rewrite  (\ref{1102}) such that it 
contains explicit projection onto it. This will complete our efforts towards 
decomposition 
of the unit operator (in the two-photon sector)  into explicit projectors onto 
all states from the physical subspace of the theory.

So, we  start calculations with 
\be
\begin{aligned}
|\Phi_2&(\k,\k')\ra=\frac{\beta}{2}L_{\k}^\dag L_{\k'}^\dag|0\ra\\
=&\frac{\beta}{\sqrt{2}}\left(
|\k3\k'3\ra  +|\k0\k'0\ra
-|\k3\k'0\ra-|\k0\k'3\ra\right),
\end{aligned}
\label{P2}
\ee
where the normalizing  prefactor $\beta$
will be determined later on 
\cite{remarkNormalization}. 
Without loss of generality, real $\beta\neq0$ will be
considered below.

Then, we  find  two states with two unphysical
photons having $\k$ and $\k'$ momenta, which  are not only mutually orthogonal but also  orthogonal to (\ref{P2})
and linearly independent from it
\begin{align}
\label{IIp}
&|\II'(\k,\k')\ra=\frac{1}{\sqrt{2}}\B{|\k3\k'3\ra-|\k0\k'0\ra},\\
&|\II''(\k,\k')\ra=\frac{1}{\sqrt{2}}\B{|\k3\k'0\ra-|\k0\k'3\ra}.
\label{IIpp}
\end{align}
Their normalization is chosen  such that the projectors
\begin{align}
&P'=\int\d{k}\d{k'} |\II'(\k,\k')\rl\II'(\k,\k')|      ,\\
&P''=\int\d{k}\d{k'}|\II''(\k,\k')\rl\II''(\k,\k')|
\end{align}
satisfy 
$P'|\II'(\p,\p')\ra=|\II'(\p,\p')\ra$
and $P''|\II''(\p,\p')\ra=-|\II''(\p,\p')\ra$.
These two states  do not belong to
$\HGB$.

Next, we find a family of states
labeled by the parameter $\alpha\in\mathbb{C}\setminus\{-1\}$, 
which  are distinct (linearly independent) from  $|\Phi_2(\k,\k')\ra$
and orthogonal to both 
$|\II'(\k,\k')\ra$ and  $|\II''(\k,\k')\ra$: 
\be
\begin{aligned}
|\II_\alpha(\k,\k')\ra=&\gamma\left(|\k3\k'3\ra
+|\k0\k'0\ra\right. \\
&\left.+\alpha|\k3\k'0\ra+\alpha|\k0\k'3\ra\right),
\end{aligned}
\label{IIalf}
\ee
where the normalizing factor $\gamma\in\mathbb{C}\setminus\{0\}$
will be  fixed in (\ref{gammam}). Such states also do not belong 
to $\HGB$.

We  use them 
to build the  projector onto  $|\Phi_2(\k,\k')\ra$, which cannot be written
as  $\int\d{k}\d{k'}|\Phi_2(\k,\k')\rl\Phi_2(\k,\k')|$
due to the zero norm of such a
state. Namely, we introduce 
\be
\begin{aligned}
P_\alpha=
\int \d{k}\d{k'} (&|\Phi_2(\k,\k')\rl\II_{\alpha}(\k,\k')| \\
+&|\II_{\alpha}(\k,\k')\rl\Phi_2(\k,\k')|)
\end{aligned}
\label{Palfa}
\ee
and  normalize (\ref{IIalf}) by choosing 
\be
\gamma=\frac{1}{\sqrt{2}\beta(1+\alpha)}
\label{gammam}
\ee
so that $P_{\alpha}|\Phi_2(\p,\p')\ra=|\Phi_2(\p,\p')\ra$.
For this to happen, one actually does
not need the second term on the right-hand side of  (\ref{Palfa}), which   is added to ensure hermiticity
of such a projector. 

By comparing   (\ref{1102}) and   $P_\alpha + P' - P''$, we find that the 
difference between the two is given by 
$\int\d{k}\d{k'}a_\alpha\beta^{-2}|\Phi_2(\k,\k')\rl\Phi_2(\k,\k')|$,
where   $a_\alpha$ has been already introduced in  (\ref{aalfa}).  

In the end, making use of the freedom to choose $\beta$, we set
\be
\beta=1
\label{beta1}
\ee
getting 
\be
\begin{aligned}
&\mathbbm{1}_{0\otimes2}=\int\d{k}\d{k'} \big(
a_\alpha|\Phi_2(\k,\k')\rl\Phi_2(\k,\k')|\\ 
&+|\Phi_2(\k,\k')\rl\II_\alpha(\k,\k')| +|\II_\alpha(\k,\k')\rl\Phi_2(\k,\k')| \\ 
&+|\II'(\k,\k')\rl\II'(\k,\k')|-|\II''(\k,\k')\rl\II''(\k,\k')|\big).
\end{aligned}
\label{02}
\ee
Two remarks are in order now.

First, thanks to (\ref{beta1}),  the 
coefficient in front of $|\Phi_2\rl\Phi_2|$ in (\ref{02}) is
the same as the one in front of  $|\Phi_1\rl\Phi_1|$ in (\ref{01new}).
Such a choice facilitates comparison between the two expressions.

Second,  the fact that (\ref{02})   is built of four linearly independent 
vectors--$|\Phi_2\ra$,
$|\II_\alpha\ra$, $|\II'\ra$, and $|\II''\ra$--reflects the number of ways 
one can distribute two unphysical photons into two momentum modes.

Looking now back at the discussion from Sec. \ref{Gauge_sec}, one can easily
check that the same scheme has been  utilized there for decomposition of unity 
in the one-photon sector.
Expressions from Sec. \ref{Gauge_sec}, however, are a bit simpler because there are  no 
 single-unphysical-photon equivalents of $|\II'\ra$
and $|\II''\ra$ that are orthogonal to $|\Phi_1\ra$.

We expect that the above-explained strategy for construction of 
(\ref{20}), (\ref{11}), and (\ref{02})
can be generalized so as to yield the unit
operator in higher photon-number sectors with explicit projectors onto  all states belonging to 
$\HGB$.
Such studies, however, are beyond the scope of this work.

To put our  results in a broader perspective, we have the following remarks.
Our   decomposition of the unit operator is built   of eigenstates of
the normal-ordered Hamiltonian describing the covariantly-quantized free
electromagnetic field  \cite{Greiner}
\be
H=\int \d{k}\om{k}(c^\dag_{\k1}c_{\k1} + c^\dag_{\k2}c_{\k2} +
c^\dag_{\k3}c_{\k3} - c^\dag_{\k0}c_{\k0}).
\ee
While working in the one- and two-photon sectors, we make use of its degenerate
eigenstates. Namely, 
$|\k\sigtr\ra$, $|\Phi_1(\k)\ra$, $|\Rzymskie{1}_\alpha(\k)\ra$ in the
one-photon sector
and 
$|\k\sigtr\k'\sigtrPrim\ra$, 
$|\Phi_2(\k,\k')\ra$, $|\II_\alpha(\k,\k')\ra$, etc. in the two-photon sector.
Degeneracy of these eigenstates follows from 
\begin{align}
&H|\k\sigma\ra=\om{k}|\k\sigma\ra, \\
&H|\k\sigma\k'\sigma'\ra=(\om{k}+\om{k'})|\k\sigma\k'\sigma'\ra,
\end{align}
which are valid for all  $\sigma,\sigma'=0,1,2,3$.

The question  how to construct the unit operator, out of eigenstates
of a hermitian operator in the indefinite-metric space, is in general quite
non-trivial \cite{Nagy,Nakanishi1972}. An algorithm for achieving 
this goal in   finite-dimensional 
systems, from eigenstates of operators having non-degenerate
spectrum, is discussed in Sec. 1.2 of \cite{Nagy}. It 
 is similar to our approach in the sense that it also involves projectors
onto zero-norm states. The details, however, are quite different 
presumably  due to the fact that we work with  degenerate eigenstates.
Taking also into account that 
our space is actually infinite dimensional, we see our results as
complementary to those reported in \cite{Nagy}.


\begin{thebibliography}{39}
\bibitem[{rem({\natexlab{a}})}]{remarkNorm}
\bibinfo{note}{For any vector $|\psi\ra$, we define the norm to be given by the
  inner product $\la\psi|\psi\ra$.}
\bibitem[{rem({\natexlab{c}})}]{remarkEqtimes}
\bibinfo{note}{All commutators in this paper, which involve two field
  operators, two angular momentum operators, or a field and an angular momentum
  operator, are computed under the equal-time assumption.}
\bibitem[{rem({\natexlab{d}})}]{remarkShell}
\bibinfo{note}{All integrals of the form $\int\d{k} f(k)$, where $f$ is an
  arbitrary function, are written under the assumption that $k^0=\om{k}$.}
\bibitem[{rem({\natexlab{e}})}]{remarkAll}
\bibinfo{note}{All expressions where some $3$-momentum is singled out, just as
  $\k$ in (\ref{GB}), are supposed to hold for all such $3$-momenta.}
\bibitem[{rem({\natexlab{g}})}]{remarkOnly}
\bibinfo{note}{The quotation marks refer to the facts that (i) (\ref{PHI1})
  actually represents a whole family of zero-norm states labeled by the
  momentum $\k$; (ii) there is no unique normalization of zero-norm states.}
\bibitem[{rem({\natexlab{i}})}]{remarkBDfield}
\bibinfo{note}{Regularization ambiguities discussed in \cite{BDfield} are
  resolved by the systematic approach developed in \cite{BDall}, which
  consistently imposes Pauli-Villars regularization across all calculations. It
  shows that ad hoc Pauli-Villars regularization attempts explored in
  \cite{BDfield} lead to the correct one-loop result for field angular momentum
  of the electron.}
\bibitem[{rem({\natexlab{j}})}]{remarkNormalization}
\bibinfo{note}{Strictly speaking, one cannot normalize a zero-norm vector such
  as $|\Phi_2\ra$. One can, however, fix the overall prefactor of such a state,
  which is what we do.}
\end{thebibliography}

\end{document}